\pdfoutput=1
\documentclass[aps,prd,twocolumn, floatfix, nofootinbib
]{revtex4}
\usepackage{amsmath}
\usepackage{amssymb}
\usepackage{epsfig}
\usepackage{graphicx}
\usepackage{color}
\usepackage{dsfont}
\usepackage{array} 

\newcommand{\be}{\begin{equation}}
\newcommand{\ee}{\end{equation}}
\newcommand{\bdm}{\begin{displaymath}}
\newcommand{\edm}{\end{displaymath}}
\newcommand{\bea}{\begin{eqnarray}}
\newcommand{\eea}{\end{eqnarray}}
\newcommand{\nn}{\nonumber}

\begin{document}

%----------------------------------------------------------------------------------
\author{Helena Kole\v{s}ov\'a}\email{helena.kolesova@fjfi.cvut.cz}
\affiliation{Institute of Particle and Nuclear Physics,
Faculty of Mathematics and Physics,
Charles University in Prague, V Hole\v{s}ovi\v{c}k\'ach 2,
180 00 Praha 8, Czech Republic \\ and \\Faculty of Nuclear Sciences and Physical Engineering, Czech Technical University in Prague, B\v{r}ehov\'a 7, 115 19 Praha 1, Czech Republic
}
\author{Michal Malinsk\'{y}\footnote{Presenting author.}}\email{malinsky@ipnp.troja.mff.cuni.cz}
\affiliation{Institute of Particle and Nuclear Physics,
Faculty of Mathematics and Physics,
Charles University in Prague, V Hole\v{s}ovi\v{c}k\'ach 2,
180 00 Praha 8, Czech Republic}
%----------------------------------------------------------------------------------

\title{Minimal $SO(10)$ grand unification at next-to-leading order}

%######################################################
\begin{abstract}
We comment on the status and prospects of the minimal non-supersymmetric renormalizable $SO(10)$ grand unified model. We emphasize its unique predictive power owing to the particular robustness of the gauge unification pattern to the leading-order Planck-scale effects which, in turn, makes it essentially the only unified framework in which the proton lifetime may be estimated with better than the leading order precision. This, together with the frequent presence of light-scale exotics accessible, in principle, at future colliders, is the key to its potential testability at the upcoming megaton-scale facilities such as Hyper-Kamiokande or LENA.  
\end{abstract}
%######################################################
\maketitle
%\begin{keyword}
%Gauge unification

%\end{keyword}

%\end{frontmatter}

% \linenumbers

\section{Introduction}
\label{sect:introduction}
%######################################################
Though being around for over forty years the idea of grand unification of electroweak and strong interactions~\cite{Georgi:1974sy} still receives significant attention, both on the theory and on the experimental sides, leaving its imprints in the steady stream of papers on baryon and/or lepton number violation and related subjects.
The activity is further boosted by the close complementarity between the modern neutrino programme and the proton decay searches which, in fact, became an important part of the trademark of essentially all the upcoming large-scale neutrino experiments such as Hyper-Kamiokande, LBNE and, possibly, LENA.  

Nevertheless, the megaton scale these facilities are about to attack makes the prospects of pushing their sensitivity to the baryon-number-violating signals much beyond the current limits rather unclear; the expected sensitivity of the Hyper-K is assumed to reach (in the principal $p\to \pi^{0}e^{+}$ channel; see Table~\ref{TabLimits} for others)
\bea
\label{HKlimit2025}\tau(p\to \pi^0 e^+)^{\rm 2030}_{{\rm HK}} &\gtrsim& 9\times 10^{34}\,\mathrm{years}\,,\\
\label{HKlimit2040}\tau(p\to \pi^0 e^+)^{\rm 2045}_{{\rm HK}} &\gtrsim& 2\times 10^{35}\,\mathrm{years}\,,
\eea
which is only about an order of magnitude higher than the current Super-Kamiokande bound 
\begin{equation}\label{SKlimit}
\tau(p\to \pi^0 e^+)_{\rm SK} > 8.2\times 10^{33}\,\mathrm{years}.
\end{equation}

\begin{table}
\begin{center}
\includegraphics[width=7cm]{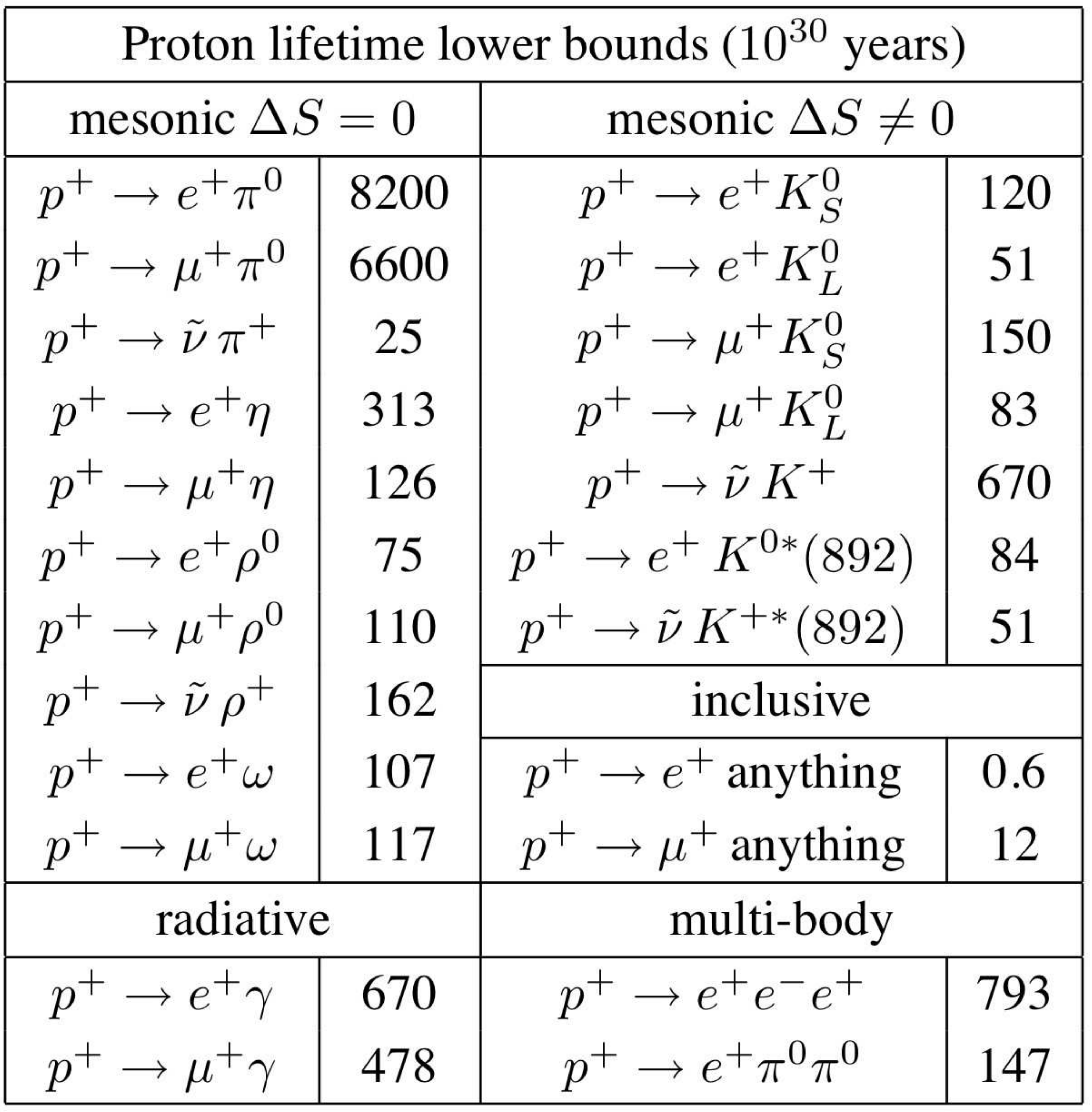}
\end{center}
\caption{The current 90\% C.L. experimental limits on proton decay in the principal two- and three-body channels.}
\label{TabLimits}
\end{table}
As the improvement rate reaches only about a factor of ten per decade and it may easily decelerate with the rocketing costs of such immense machines, the community may really benefit from such enormous investments if and only if the theory error is contained within the anticipated one-order-of-magnitude improvement window.
%######################################################
\section{Proton lifetime estimates}
\label{sect:estimates}
%######################################################
Unfortunately, the real situation is very far from this ideal as the existing estimates are extremely coarse stretching, in most cases, over several orders of magnitude, cf. Table~\ref{TabEstimates}. The main reason for this  disparity consists in the fact that, despite continuing efforts, we still have a relatively poor control over the main ingredients entering the proton lifetime estimates in grand unifications (GUTs). These are, namely:
\begin{table}
\begin{center}
\includegraphics[width=8cm]{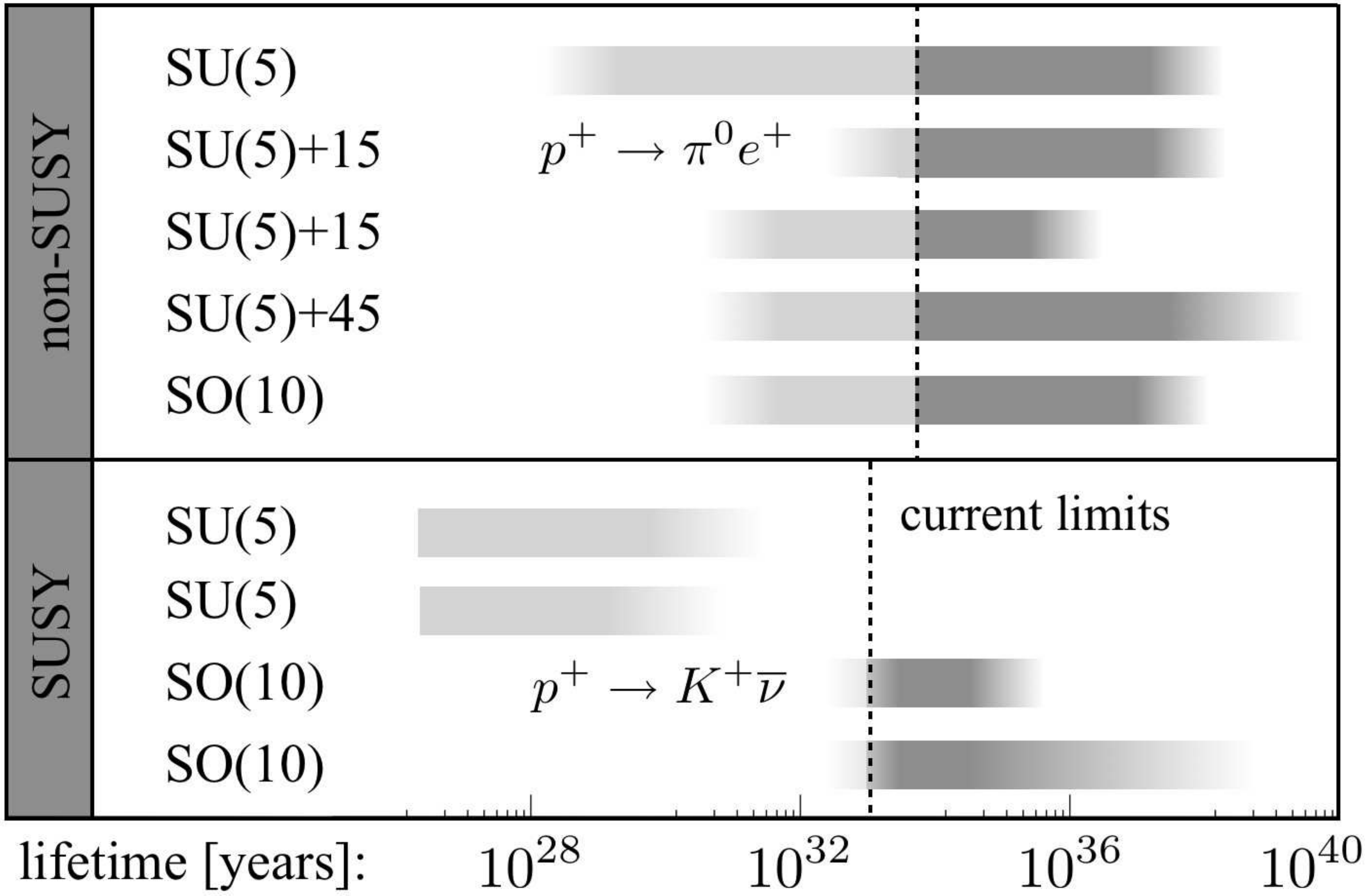}
\end{center}
\caption{A sample of the leading-order proton lifetime estimates available in the literature. The bands have been taken, consecutively, from the references  \cite{Georgi:1974yf,Goldman:1980ah}, \cite{Dorsner:2005fq}, \cite{Dorsner:2006hw}, \cite{Dorsner:2006dj}, \cite{Lee:1994vp}, \cite{Pati:2005yp}, \cite{Murayama:2001ur}, \cite{Pati:2005yp} and \cite{Dutta:2004zh}. It is clear that the indicated error bars exceed the expected ``improvement window'' of the future megaton-scale facilities by many orders of magnitude and, hence, their data may not be used for any conclusive discrimination among different scenarios.}
\label{TabEstimates}
\end{table}

{\em i) The masses of the lightest baryon-number-violating (BNV) messenger fields, typically, the relevant GUT-scale vector bosons.} This, in non-supersymmetric (non-SUSY) GUTs, enters the proton lifetime in fourth power and, hence, becomes one of the parameters of central interest. Since it is determined from the analysis of the renormalization group evolution of the low-energy effective gauge couplings which, in the vicinity of the GUT scale, typically intersect under relatively shallow angles (and, moreover, only as logs of the renormalization scale), the corresponding experimental and theoretical uncertainties tend to be amplified enormously.

{\em ii) The flavour structure of the BNV currents.} With the relatively limited information about the flavour structure of the Standard model (especially the individual Yukawa couplings) this is usually rather difficult to constrain to such a level to draw any definite and detailed conclusions. Although the related uncertainties tend to propagate only polynomially (rather than exponentially) into the proton lifetime estimates a detailed account of the relevant effects requires a thorough analysis of the Yukawa-sector matching conditions. Since, however, these are generally strongly model dependent, a vital ingredient of any analysis of this kind is a delicate balance between the need to accommodate all existing flavour data on one side and the often overwhelming complexity of the model's Higgs sector on the other side.

{\em iii) The hadronic matrix elements relating the d=6 quark-level hard processes to the hadronic-level observables.} Until recently, these amounted to another serious source of theoretical uncertainties plaguing all the attempted proton lifetime estimates. Luckily, a steady progress on the lattice and chiral perturbation theory methods brought these quantities to such a level of  theoretical understanding (amounting to relative errors of the order of  several tens of percent) that they are no longer the primary concern for the accuracy of the proton lifetime estimates. 

This (incomplete) list makes it also clear why the desired level of accuracy has never been attained. As for the ``prehistory'' of the subject (until about mid 1980's) there were no good data on the low-scale gauge couplings so most of the early attempts were more or less academic. By their final arrival, the simplest Georgi-Glashow $SU(5)$ model~\cite{Georgi:1974sy} was shown to be in trouble due to its overly large weak mixing angle post-diction. In parallel, the same data was reclaimed to be rather a strong hint about the low-energy SUSY so the failure of the simplest non-SUSY GUTs did not bother too much. However, the lack of any solid information about the SUSY spectrum, with its obvious impact on the accurate proton lifetime predictivity prospects, inhibited most of the attempts to tame the relevant theoretical uncertainties for almost two decades. The short renaissance of the SUSY $SO(10)$ GUTs at the beginning of the last decade, owing to the feesch leptonic flavour data supplied by the discovery of neutrino oscillations, left behind a yet deeper depression after the minimal such models were shown to be all sick in one way or another \cite{Bertolini:2006pe,Aulakh:2005mw}.

As a matter of fact, even if things did not take such a desperate turn for the minimal GUTs, there was only a little perspective in trying to transform the vastly improved inputs into a good quality proton lifetime estimate at the next-to-leading order (NLO; this, however, is the minimum level of perturbation expansion suitable for addressing the aforementioned accuracy issues). The reason, as always, is in the proximity of the Planck scale ($M_{Pl}$) which does not admit ``sweeping the gravity-induced non-renormalizable operators under the carpet''.  The most ruinous of these are usually the d=5 corrections to the gauge-field kinetic form emerging from the operator of the kind
\be\label{d5dangerous}
{\cal L}\ni \frac{C}{M_{Pl}}G^{a}_{\mu\nu}\Phi^{ab}G^{b\mu\nu}
\ee
where $G_{\mu\nu}^{a}$ is the gauge-field tensor, $\Phi$ is the GUT symmetry breaking Higgs field; indeed, such a structure in the asymmetric phase yields an irreducible theoretical uncertainty in the normalization of gauge fields and, thus, in the definition of the effective gauge couplings by means of the GUT-scale matching conditions. Needless to say, due to the logarithmic nature of the renormalization group running such (typically per-cent level) effects tend to ruin totally the accuracy of the GUT-scale determination unless the unknown $C$ coefficient is strongly suppressed. Moreover, the rich high-energy structure of namely the SUSY GUTs tends to lower the effective cut-off of of such models by more than an order of magnitude below the (reduced) Planck scale (see, e.g.,~\cite{Dvali:2007hz}), which only further adds to the overall trouble.  

Hence, without at least some control over the operators of the kind (\ref{d5dangerous}) there is no point in even attempting to go beyond the leading order approximation in GUT estimates of the proton lifetime, let alone achieving the required accuracy goals.  
From this perspective, there is practically a unique scenario among the classical $SU(5)$ and/or $SO(10)$ GUTs in which all the aforementioned issues may be efficiently dealt with, namely, the non-SUSY $SO(10)$ model in which the GUT-scale symmetry breaking is triggered by an adjoint scalar. The point is that, unlike with the other options, the full antisymmetry  of the 45-dimensional tensor makes the most dangerous gravity-induced operator (\ref{d5dangerous}) vanish and, thus, avoids the most serious obstacle on the quest for the NLO proton lifetime estimates. 

%######################################################
\section{The Minimal non-SUSY $SO(10)$ GUT}

%######################################################
\subsection{Prehistory}
\label{sect:minimalSO10prehistory}
Unlike in the $SU(5)$ case where there is only one symmetry breaking pattern $SU(5)\to SU(3)c\otimes SU(2)_{L}\otimes U(1)_{Y}$ to be considered (pinning down only a couple of reasonable options for the Higgs mechanism implementation at the perturbative and renormalizable level) the higher rank of the $SO(10)$ gauge group makes a systematic approach to the $SO(10)$ unifications much more involved. Indeed, there is more than a dozen different subgroups of $SO(10)$ available as effective intermediate-scale symmetries compatible with the Standard model (SM)~\cite{Deshpande:1992au} out of which almost all may be reached in more than one way within different Higgs models (even at the renormalizable level). From this perspective, every argument constraining or discarding at least some of the options is  welcome, and it was even more so  in the early days. For instance, at the beginning of 1980's the issue of the monopole overproduction in the early Universe was taken~\cite{Lazarides:1980cc,Harvey:1981hk} as a rationale to prefer the models with the scalar 54 triggering the GUT-scale symmetry breaking over those in which the same was achieved by an adjoint 45 instead. Although this line of argument was invalidated after the invention of the cosmic inflation in 1982~\cite{Guth:1980zm} the models with 45 in the Higgs sector did not enjoy much of a renaissance due to another pathology having to do with a proliferation of tachyonic instabilities along all the potentially realistic symmetry breaking chains.   
\subsection{The trouble with the adjoint $SO(10)$ Higgs model}
\label{sect:thetrouble}

The point is that the classical scalar potential 
\be
\label{scalpotgen}
V = V_{45} + V_{126} + V_{\rm mix} \, ,
\ee
where (see, e.g., ~\cite{Bertolini:2012im} for the notation details)
\bea
\label{V45}
V_{45} &=& - \frac{\mu^2}{2} (\phi \phi)_0 +\!\frac{a_0}{4} (\phi \phi)_0 (\phi \phi)_0\\ &+&\! \frac{a_2}{4} (\phi \phi)_2 (\phi \phi)_2, \nn \\
\label{V126}V_{126} &=&  - \frac{\nu^2}{5!} (\Sigma \Sigma^*)_0+\frac{\lambda_0}{(5!)^2} (\Sigma \Sigma^*)_0 (\Sigma \Sigma^*)_0\\
& +& 
\frac{\lambda_2}{(4!)^2} (\Sigma \Sigma^*)_2 (\Sigma \Sigma^*)_2\nn\\
& + & \frac{\lambda_4}{(3!)^2(2!)^2} (\Sigma \Sigma^*)_4 (\Sigma \Sigma^*)_4\nn\\
&+ &\frac{\lambda'_{4}}{(3!)^2} (\Sigma \Sigma^*)_{4'} (\Sigma \Sigma^*)_{4'} \nn \\ \nn \\
&+& \left[\frac{\eta_2}{(4!)^2} (\Sigma \Sigma)_2 (\Sigma \Sigma)_2
+ h.c.\right] \, , \nn \\ \nn \\
\label{V45126}
V_{\rm mix} &=& \frac{i \tau}{4!} (\phi)_2 (\Sigma \Sigma^*)_2
+ \frac{\alpha}{2 \cdot 5!} (\phi \phi)_0 (\Sigma \Sigma^*)_0 \\
&+& \frac{\beta_4}{4 \cdot 3!} (\phi \phi)_4 (\Sigma \Sigma^*)_4
+ \frac{\beta'_{4}}{3!} (\phi \phi)_{4'} (\Sigma \Sigma^*)_{4'} \nn \\ \nn \\
&+& \left[\frac{\gamma_2}{4!} (\phi \phi)_2 (\Sigma \Sigma)_2
+h.c.\right] \,\nn,
\eea
yields an overly strong constraint on the SM-compliant VEV direction of the adjoint Higgs field 
$$
\langle \Phi \rangle = {\rm diag}\{\omega_{BL},\omega_{BL},\omega_{BL},\omega_{R},\omega_{R}\}\otimes
\left(\begin{array}{cc}
0 & 1\\
-1 & 0
\end{array}\right)
$$ 
due to the resulting tachyonicity of the pair of the scalar masses (classified by the SM quantum numbers)
\bea
\label{treemass130}
M^2_{(1,3,0)} &=& \!2 a_2 (\omega_{BL}\! -\! \omega_R) (\omega_{BL} + 2 \omega_R) \, , \\
\label{treemass810}
M^2 _{(8,1,0)} &=& \!2 a_2 (\omega_R\! -\! \omega_{BL}) (\omega_R + 2 \omega_{BL}) \,,
\eea
outside the 
\be\label{treecondition}
a_{2}<0 \,,\quad -2 < \frac{\omega_{BL}}{\omega_{R}}<-\frac{1}{2} 
\ee
domain. Indeed, for $\omega_{BL}$ close to $\omega_{R}$ the setting is too close to an exact $SU(5)\otimes U(1)$ limit ($\omega_{BL}=-\omega_{R}$) which, however, is incompatible with the non-SUSY gauge unification constraints.
%######################################################
\subsection{The quantum resurrection of the minimal $SO(10)$ model}
\label{sect:minimalSO10ressurection}
%######################################################
However, a dedicated 2009 analysis of the adjoint $SO(10)$ Higgs model~\cite{Bertolini:2009es} revealed  that this pathology was actually an artefact of the tree-level approximation and that it could be overcome at the quantum level. On the technical level this owes to a set of large positive radiative corrections to the potentially tachyonic masses (for instance  
\bea
\label{310onthevac}
\Delta M^2_{(1,3,0)}\!\!&=&\!\!\frac{g^4}{4\pi^2}
 \left(16 \omega _R^2+\omega _{BL} \omega _R+19 \omega _{BL}^2\right)+\ldots,\nn\\[1ex]
\label{810onthevac}
\Delta M^2_{(8,1,0)}\!\!&=& \!\!\frac{g^4 }{4\pi^2} 
\left(13 \omega _R^2+\omega _{BL} \omega _R+22 \omega _{BL}^2\right)+\ldots \nn,
\eea
from the gauge loops) that may easily overshoot the problematic tree-level contributions (\ref{treemass130}) and (\ref{treemass810}). Thus, the tachyons may be tamed also outside the (\ref{treecondition}) domain and the minimal renormalizable $SO(10)$ Higgs model was brought back from oblivion.%######################################################
\section{The minimal $SO(10)$ unification at next-to-leading order}
\label{sect:minimalSO10now}
%######################################################
The next obvious question to be addressed is whether the adjoint  $SO(10)$ Higgs mechanism may support a full-fledged grand unified model. As a matter of fact, the preceding analyses of the symmetry breaking patterns, e.g., \cite{Bertolini:2009qj}  based on the minimal survival hypothesis~\cite{delAguila:1980at} (and, hence, ignoring the details of the scalar spectrum) indicated that the scale at which the ${B-L}$ gauge symmetry is broken is somewhat on the low side, typically in the $10^{10-11}$ GeV ballpark. This, however, turns out to be problematic for the implementation of the seesaw mechanism regardless of whether the $B-L$ breaking is triggered by the VEV ($\sigma$) of a 16-dimensional spinor or a 126-dimensional (self-dual part of the) fully antisymmetric 5-index tensor\footnote{Let us note that even in the former case where there is no suitable  renormalizable Yukawa coupling to give rise to the RH neutrino masses at the tree level these may nevertheless be generated at two loops by the Witten's mechanism~\cite{Witten:1979nr}.}. 
However, even in the latter case (for which the seesaw scale depends linearly on $\sigma$ and, hence, it is not as suppressed as in the former scenario where the same dependence is quadratic) the ``naive'' $\sigma \lesssim 10^{11}$ GeV bound makes it possible to accommodate the light neutrino masses only for the price of a multiple fine-tuning in the neutrino Dirac mass formula. 

\subsection{Minimally fine-tuned settings}
However, there is a much more economic way to get the light neutrino mass scale right without disturbing the basic consistency requirement of gauge unification -- it is sufficient to trade the single fine-tuning (fixing $\sigma\sim 10^{11}$ GeV) in the one-point Green's function of the $B-L$ breaking scalar for an alternative fine-tuning in a two-point function (i.e., mass) of another suitably chosen scalar. Strictly speaking, this class of options should be viewed as no worse than the ``traditional'' ones because the number of fine-tunings remains the same.       

\subsection{Basic constraints}
Of course, in doing so one, has to conform a number of consistency constraints, namely:

{\em i) Perturbativity.} This is clearly a vital ingredient of any potentially accurate perturbative account. 

{\em ii) Vacuum stability.} We shall require that any potentially realistic choice of the model's parameters, in particular, of the VEVs $\omega_{BL}$ and $\omega_{R}$, is such that the scalar spectrum is non-tachyonic and, hence, the vacuum is at least metastable. Since, however, the classical scalar spectrum tends to be tachyonic, cf. (\ref{treemass130})-(\ref{treecondition}), this question may be fully addressed only at the quantum level. Unfortunately, with the current state of the art the knowledge of the scalar sector quantum structure in the 45+126 model is restricted to just a handful of leading order polynomial corrections; nevertheless, these should be enough~\cite{Bertolini:2009es} to get a good grip on the allowed vacuum configurations even without a complete one-loop effective-potential-based calculation. 
  
{\em iii) Unification @ NLO.} This, together with the vacuum stability requirement, turns out to be the main discriminator among the possible singly fine-tuned scenarios with different ``accidental thresholds'' in the GUT desert. The need to go beyond the LO approximation is again motivated by the prospects to estimate the proton lifetime with unprecedented accuracy.   

{\em iv) Proton decay.} Needless to say, one should conform all the current limits in Table~\ref{TabLimits}.  Focusing on the principal channel the relevant NLO formula reads
\bea\label{Gamma}
 \Gamma(p\to \pi^0 e^+)=\frac{\pi\, m_p\,\alpha _G^2 }{4
   f_{\pi}^2}  |\alpha|^2 A_L^2 (D+F+1)^2  \qquad \mbox{}\nn\\
\times\left(A_{SR}^2 \left(\frac{1}{M_{(X',Y')}^2}+\frac{1}{M_{(X,Y)}^2}\right)^2 + \frac{4 A_{SL}^2}{M_{(X,Y)}^4}\right)\qquad \mbox{}\nn
\eea
where $M_{(X,Y)}$ and $M_{(X',Y')}$ are the masses of the relevant GUT-scale vector mediators of the specific $d=6$ BNV transitions, $\alpha$, $D$ and $F$ are chiral form-factors related to the transition from the quark- to the hadronic level and $A_{L}$, $A_{SL}$ and $A_{SR}$ are renormalization factors carrying the information about the running of the effective $d=6$ operators from the GUT scale where they emerge down to the sub-electroweak scale of the very process; for further details an interested reader is deferred to the classical monography \cite{Nath:2006ut}.  
\subsection{Consistent minimally fine-tuned settings}
Remarkably enough, there turn out to be only two~\cite{Bertolini:2012im} minimally fine-tuned settings that conform all the requirements above and, at the same time, admit the $B-L$ scale in the $10^{12-14}$ GeV range generally favoured by the neutrino data.  
\subsubsection{The TeV-scale octet scenario}
\begin{figure}[h]
\begin{center}
\includegraphics[width=7.6cm]{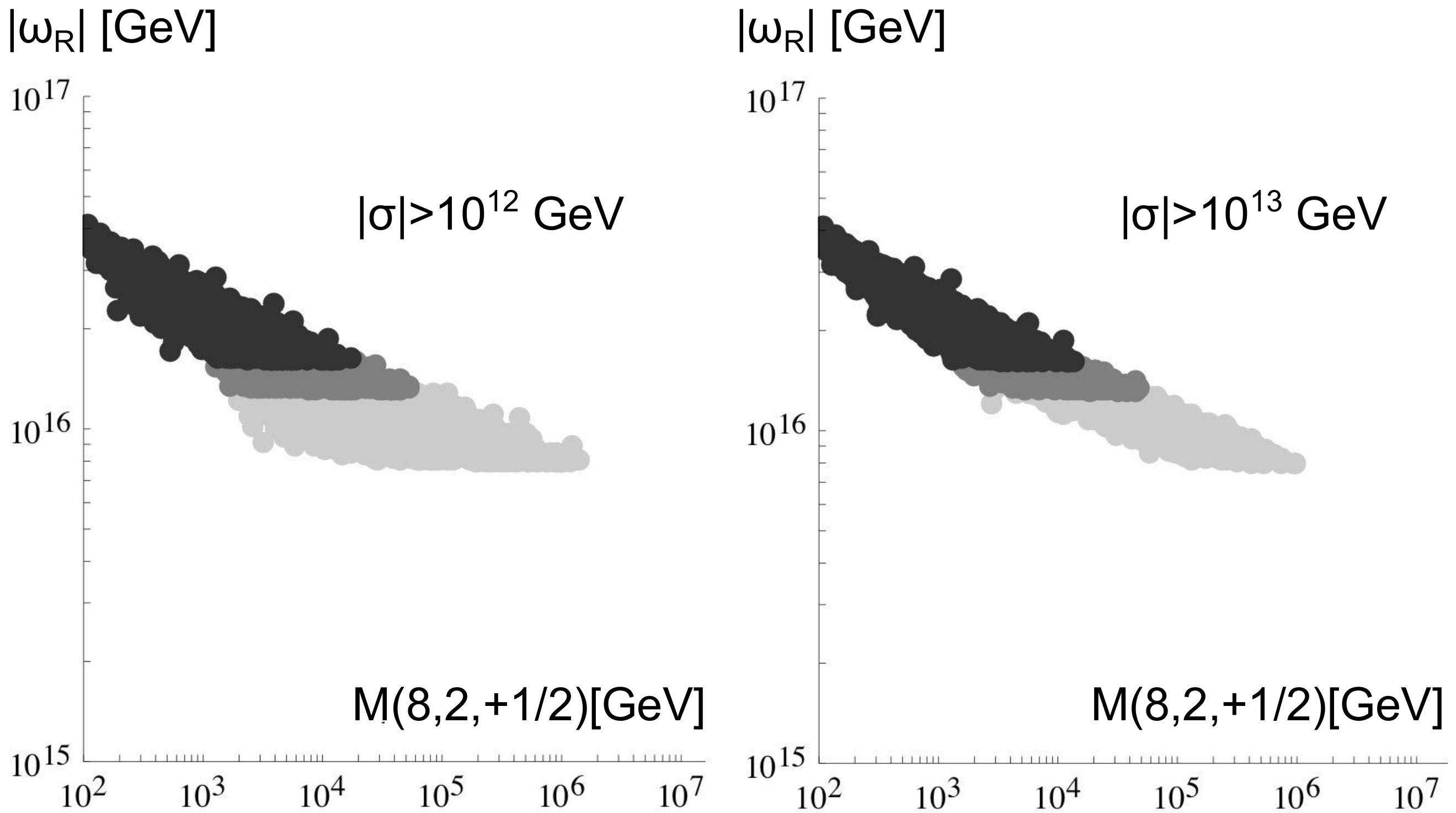}
\end{center}
\caption{The TeV-scale octet solution at the next-to-leading order.}
\label{FigOctet}
\end{figure}

The first of these is the setting featuring an accidentally light coloured octet transforming as $(8,2,+\frac{1}{2})$ under the SM gauge group, cf.~\cite{Bertolini:2013vta}. This situation is very interesting due to a strong anticorrelation between the mass of such a state and the proton lifetime depicted in Figure~\ref{FigOctet}. If, for instance, proton decay would not be seen at Hyper-K, the allowed octet mass range is fully contained under about 20 TeV which, hopefully, should be within the reach of the next-generation LHC-type hadron colliders. 

\subsubsection{The ZeV-scale sextet scenario}
\begin{figure}[h]
\begin{center}
\includegraphics[width=7.6cm]{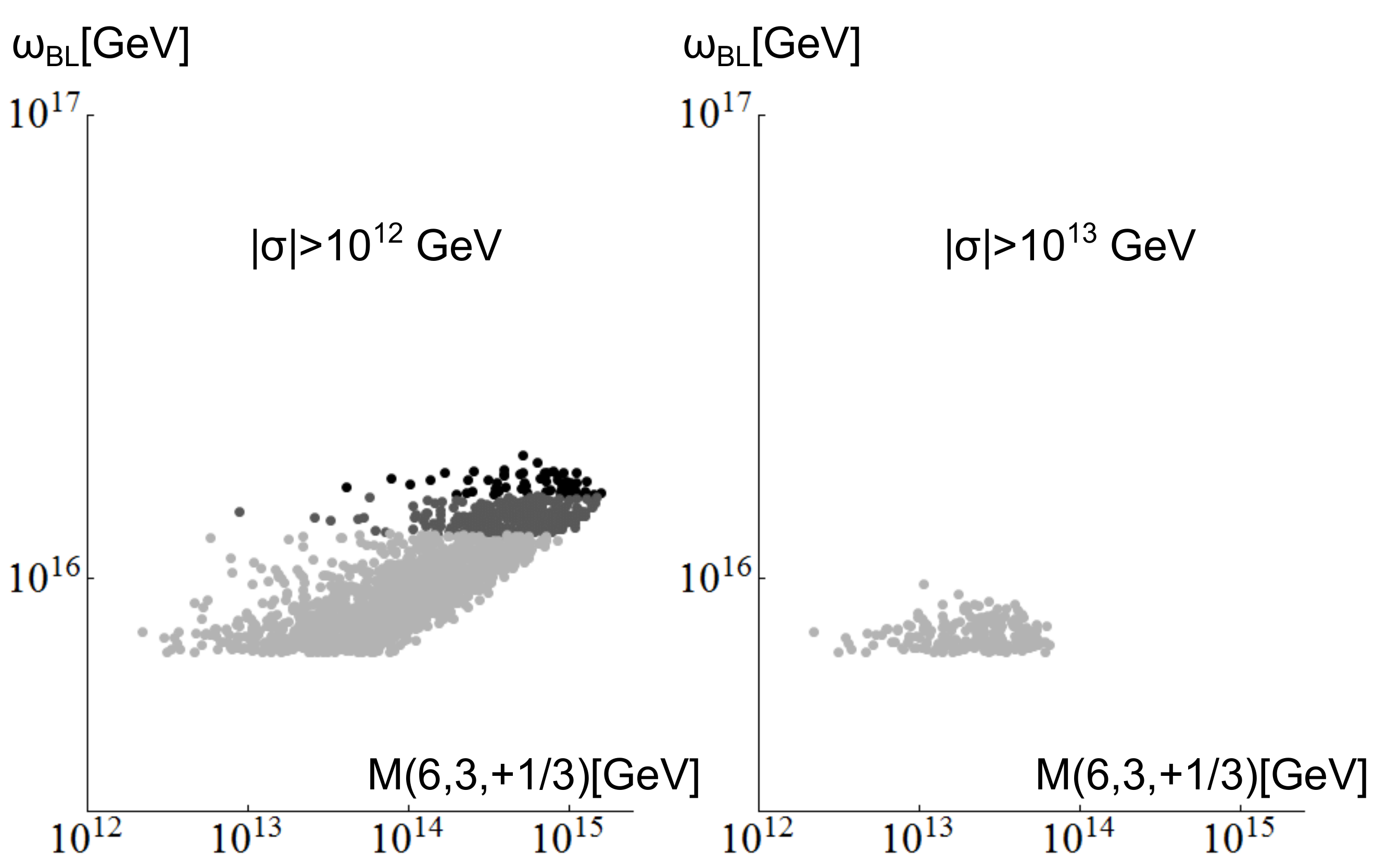}
\end{center}
\caption{The ZeV-scale sextet solution at the next-to-leading order.}
\label{FigSextet}
\end{figure}

The second potentially realistic (yet minimally fine-tuned) scenario features a scalar coloured sextet transforming  as  $(6,3,+\frac{1}{3})$ under the SM gauge group~\cite{Kolesova:2014mfa}. Though, maybe, not as attractive as the octet option, this setting is not entirely uninteresting either because it may still be refutable, see Figure~\ref{FigSextet}. Indeed, non-observation of proton decay at Hyper-K would, in this case, impose a rather strict upper bound on the seesaw scale which, within a the resulting seesaw model, may be incompatible with the absolute neutrino masses scale.   
%######################################################
\section{Conclusions and outlook}
\label{sect:conclusions}
%######################################################
The take-home message of this overview is primarily the fact that there indeed exists a simple and potentially realistic grand unified scenario in which one of the principal theoretical uncertainties plaguing traditionally the accuracy of the proton lifetime estimates (namely, the gravity smearing emerging usually at the level of $d=5$ Planck-scale induced effective operators), may be entirely under control. 

Interestingly, this scenario (that we dare to call the minimal $SO(10)$ GUT) has been slipping the attention of the community for almost two decades, partly because of its out-of-fashion non-SUSY nature and partly due to an early tree-level no-go argument; however, with the recent developments both these prejudices have been reassessed and the model was resurrected as a viable and even potentially testable physical scenario. In passing, two particularly interesting (yet minimally fine-tuned) scenarios with accidentally light states\footnote{As a mere curiosity, let us mention that at the same time the same fields have been independently singled out in the study~\cite{Dorsner:2012pp} as possible agents behind then observed excess of two-gamma events in the Higgs searches at the LHC.} in the GUT desert were identified, each of which may be even testable at the near future experimental facilities.

Nevertheless, a decisive assessment of these results is clearly premature as there are still many open issues to be addressed. Besides the necessity to resolve the technical demands of the full-fledged analysis of the one-loop effective potential in the 45+126 scenario there is namely the urgent need to include the constraints from the low-energy flavour data that, as a matter of fact, may  be the most complicated part of the business because of the non-uniqueness of the definition of the Yukawa sector in even the simplest non-SUSY $SO(10)$ GUTs~\cite{Bajc:2005zf}.  

%######################################################
\section*{Acknowledgments}
%######################################################
The work of M.M. is supported by the Marie-Curie Career Integration Grant within the 7th European Community Framework Programme
FP7-PEOPLE-2011-CIG, contract number PCIG10-GA-2011-303565, by the Charles University grant PRVOUK P45, by the Foundation for support of science and research ``Neuron'' and by the Research proposal MSM0021620859 of the Ministry of Education, Youth and Sports of the Czech Republic. The work of H.K. is supported by the Grant Agency of the Czech Technical University in Prague, grant No. SGS13/217/OHK4/3T/14.

%\bibliographystyle{h-physrev5}
%\bibliography{ICHEPbibliography}

\end{document}